\documentclass[a4paper,11pt]{article}
\pdfoutput=1 % if your are submitting a pdflatex (i.e. if you have
\usepackage{jinstpub} % for details on the use of the package, please
\usepackage{lineno}
\newcommand{\enlaser}{{\mathcal E}_\textrm{L}}
\newcommand{\ecrit}{{\mathcal E}_\textrm{cr}}

\title{\boldmath Studies of high-field QED with the LUXE experiment at the European XFEL}

\author[a,b,1]{M. Borysova,\note{Corresponding author .}}
\affiliation[a]{Deutsches Elektronen-Synchrotron DESY, Germany}
\affiliation[b]{Institute for Nuclear Research NASU (KINR), Kyiv, 03680, Ukraine}
\emailAdd{maryna.borysova@desy.de}

\abstract{The LUXE experiment aims at studying high-field QED in electron-laser and photon-laser interactions, with the 16.5 GeV electron beam of the European XFEL and a laser beam with power of up to 350 TW. The experiment will measure the spectra of electrons and photons in non-linear Compton scattering where production rates in excess of $10^9$ are expected per 1 Hz bunch crossing. At the same time positrons from pair creation in either the two-step trident process or the Breit--Wheeler process will be measured, where the expected rates range from $10^{-3}$ to $10^4$ per bunch crossing, depending on the laser power and focus. These measurements have to be performed in the presence of low-energy high radiation-background. To meet these challenges, for high-rate electron and photon fluxes, the experiment will use Cherenkov radiation detectors, scintillator screens, sapphire sensors as well as lead-glass monitors for back-scattering off the beam-dump. A four-layer silicon-pixel tracker and a compact sampling electromagnetic calorimeter will be used to measure the positron spectra. The layout of the experiment and the expected performance under the harsh radiation conditions will be presented.}

\keywords{Large detector systems for particle and astroparticle physics, Detector modelling and simulations}

\collaboration[c]{on behalf of LUXE}

\proceeding{22$^{\text{d}}$ International Workshop on Radiation Imaging Detectors
\\
  27 June 2021 - 1 July 2021\\
  online}

\begin{document}
\maketitle
\flushbottom

\section{Introduction}
\label{sec:intro}

\begin{figure}[ht]
    \begin{center}
    \includegraphics[width=0.6\textwidth]{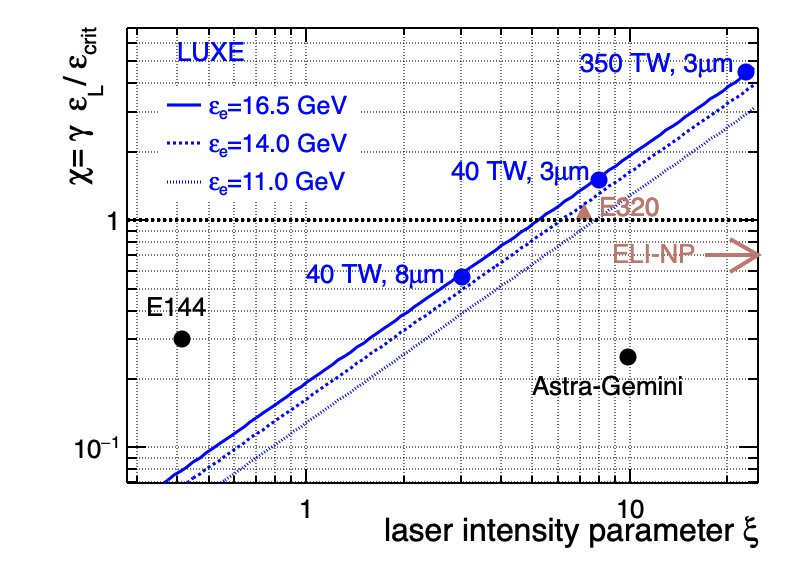}
    \caption{Quantum parameter $\chi$ vs the intensity parameter $\xi$ for a selection of experiments and facilities (E144~Ref.~\cite{Bamber:1999zt}, E320~Ref.~\cite{E320}, and Astra-Gemini~Ref.~\cite{Poder:2018ifi}). The regime of ELI-NP is also indicated. ELI-NP and E320 are not yet operating while E144 and Astra-Gemini have already published results. For LUXE, three beam energies are shown as isolines, and two laser focus spot sizes are highlighted for the 40 TW laser and one for the 350 TW laser. 
}
      \label{fig:chivsxi}
    \end{center}
\end{figure}

The LUXE experiment is a new experiment proposed at DESY and Eu.XFEL in Hamburg~\cite{luxecdr, luxeloi}. Via collisions of the XFEL electron beam and a high-power laser, LUXE will study non-perturbative and non-linear QED phenomena in the strong-field regime.
Due to the availability of high-intensity lasers based on chirped pulse amplification, LUXE can study the most prominent phenomenon -- the production of electron-positron pairs by field-induced tunnelling out of the vacuum, also called Schwinger pair production~\cite{Schwinger:1951nm}. For the field strengths well below critical $\ecrit=m_e^2c^3/(e\hslash) = 1.32 \cdot 10^{18}~\mathrm{V/m}$, 
%$1.3 \cdot10^{20}~W/cm^2$,
%$\epsilon_{cr} =1.3*10^{20}~W/cm^2$
this process is exponentially suppressed and thus basically immeasurable. Only in the strong-field regime may Schwinger pair production be observed, and this would represent a remarkable achievement.
LUXE is planning to investigate strong field QED in a yet to be explored parameter regime as can be seen in figure~\ref{fig:chivsxi}. 
There are two dimensionless parameters characterising such interactions: first, the intensity of the laser field,  $\xi = \frac{e\enlaser}{m_e\omega}$
where $\omega$ is frequency of the laser, $\enlaser$ - the field strength in the laboratory frame and $e, m_e$, the charge and mass of electron, respectively. The region $\xi \ll 1$ corresponds to the perturbative regime. The other
key parameter is the quantum non-linearity parameter $\chi$. For electron-laser interactions, $\chi = \frac{E_*}{\ecrit}$, where $E_{*}$ is the field strength in the rest frame of electron.
A unique feature of the LUXE experiment is the combination of high intensity laser parameter $\xi$ and a high electron beam energy to reach high values of $\chi$. 
For the initial phase (phase-0) of the experiment an already available 40 TW Ti:Sp laser will be used, achieving intensities up to ~$10^{20}~\mathrm{W/cm^2}$.  In a second phase (phase-1) of the experiment a more powerful laser of 350 TW is foreseen to reach intensities exceeding $10^{21}~\mathrm{W/cm^2}$.

%There are two dimensionless parameters characterising such interactions: first, $\xi$, it represents the work performed by the field over the wavelength expressed in units of electron mass  and essentially is the intensity of the laser. 
The LUXE experimental program will be accomplished by using the high-quality electron beam and a high-power laser with complex diagnostics as well as a powerful detection system. In this paper the detector technologies chosen for the various regions of the experimental setup are discussed focusing on the feasibility of the measurements.
      
\section{Physics processes} 
\begin{figure}[htbp]
\centering
\includegraphics [width=0.48\textwidth]{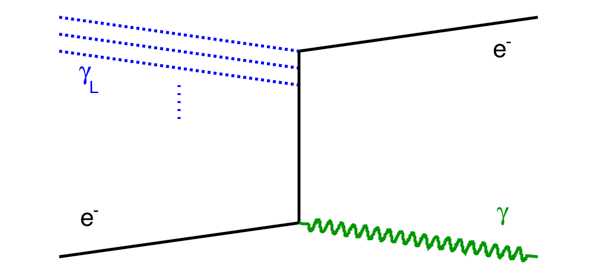} 
\includegraphics [width=0.48\textwidth]{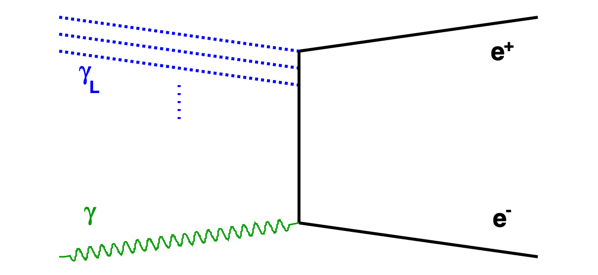} 
\caption{Schematic diagrams for the non-linear Compton process ($e^{-} + n\gamma_L \to e^{-} + \gamma$) and the non-linear Breit--Wheeler process ($\gamma + n\gamma_L \to e^+e^-$).} 
\label{fig:feyn} 
\end{figure}

The LUXE experiment will focus on the non-linear processes of Compton scattering, Breit--Wheeler pair production and trident pair production.  The Feynman diagram of the non-linear Compton scattering process $e^{-} + n\gamma_L \to e^{-} + \gamma$ is presented in figure~\ref{fig:feyn}, (left). The non-linearity is shown as possibility for one electron to interact with many laser photons to generate in final state high energy photon. Another process to investigate is Breit--Wheeler pair-creation observed in interaction of gamma rays with laser photons $\gamma + n\gamma_L \to e^+e^-$. The Feynman diagram of this process is shown in figure~\ref{fig:feyn} (right).

If the photon generated in non-linear Compton scattering~(figure \ref{fig:feyn}, left)
interacts with the laser field during the same collision and produces an electron
positron pair as in diagram (figure \ref{fig:feyn}, right), the process is referred to
trident process. The intermediate photon can also be virtual. As
follows from the diagrams, the particles in the final states are photons,
electrons and positrons, with the last ones indicating pair creation in
light-light interaction. The rate of the final state particles ranges from
$10^{-3}$ up to $10^{9}$ which imposes challenges in detector design. 
%~\cite{Hartin:2018sha}.
The custom-built strong-field QED Monte Carlo (MC) computer code PTARMIGAN \cite{ptarmigan} is used to simulate these strong field interactions for LUXE processes.

\begin{figure}[htbp]
\centering
\includegraphics [width=0.48\textwidth]{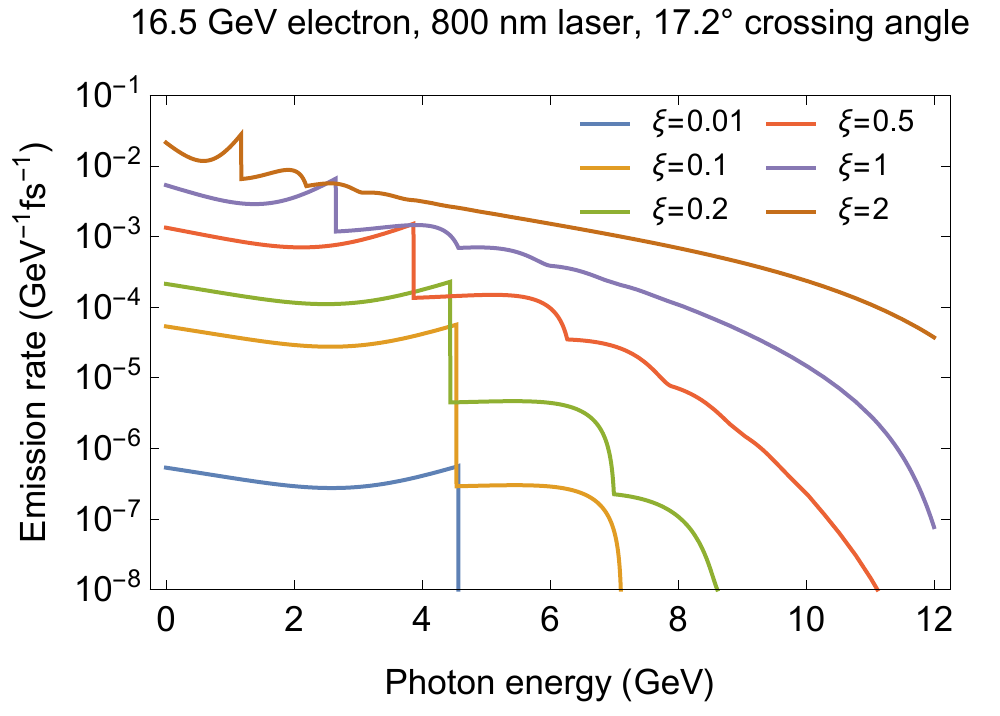} 
\includegraphics[width=0.48\textwidth]{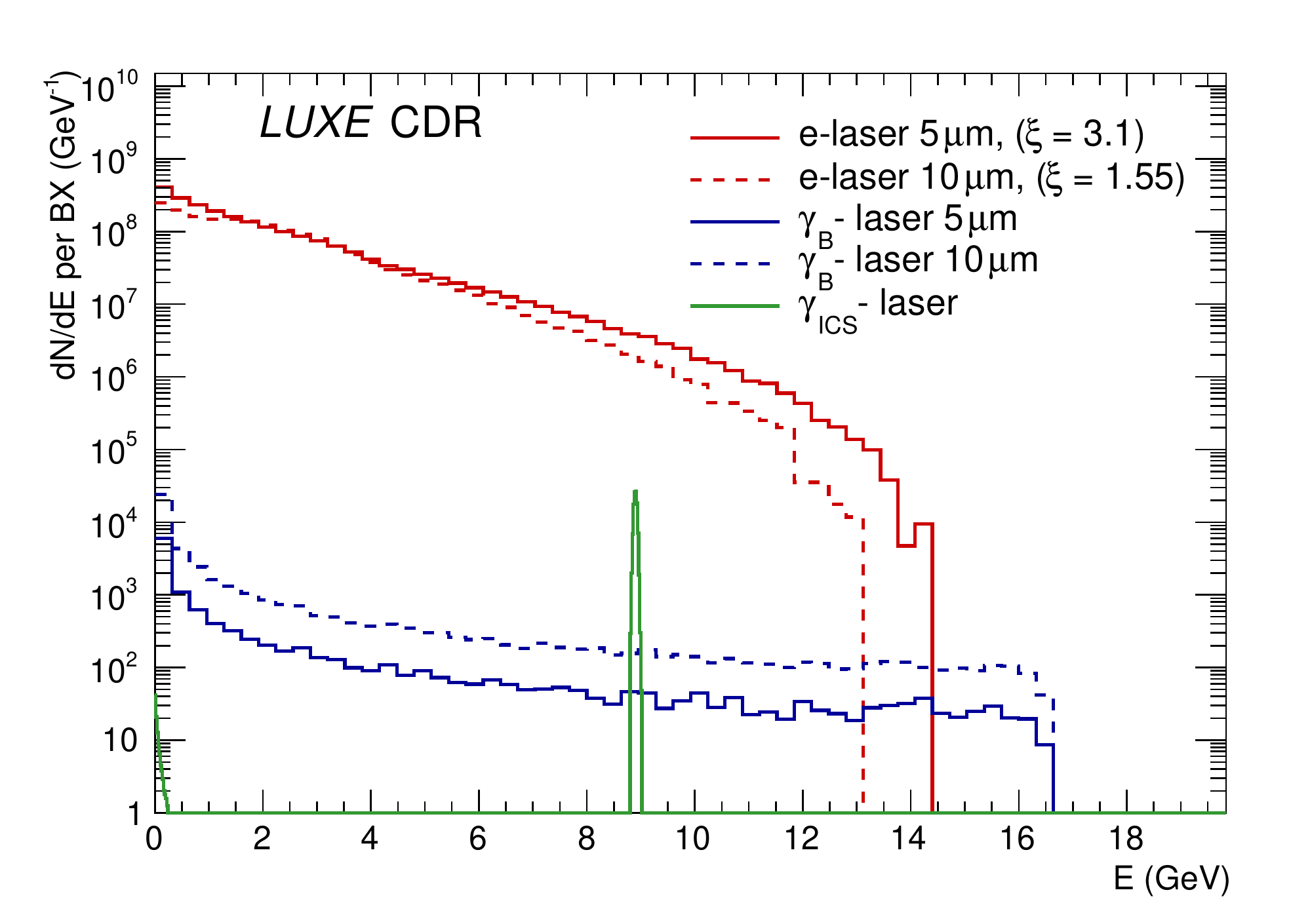}
\caption{(left) The photon emission rate for the Compton process as function of the photon energy, based on MC calculations for a selection of $\xi$ values; 
(right) Photon energy spectra for the Compton, bremsstrahlung and ICS processes for two different laser spot sizes $w_0$. For the ICS process the $w_0$ used here is $8~\mu m$. Shown are the numbers of photons within the given laser spot size.}
\label{fig:sim:hics}
\end{figure}

The photon energy spectrum for different fixed laser intensities (figure~\ref{fig:sim:hics}, left) exhibits kinematic edges, typical for Compton photon spectra. As the intensity increases the edges shift towards lower energies and contributions from multi-photon interactions get larger. The shift of the kinematic edge is explained by an increase of the effective electron mass in the strong field $m_{*}^{2} = m_e^{2}(1+\xi^2)$. The position of the edges and the intensity of multi-photon interactions are the main observables for the LUXE experiment.

For the non-linear Breit--Wheeler process the initial high energy photon can be produced via bremsstrahlung or in inverse Compton scattering (ICS) using an additional laser beam.
The plot to the right in figure~\ref{fig:sim:hics} shows photon spectra for all three possibilities including the trident process.
The bremsstrahlung setup can generate photons with the highest energies but relatively low intensity.
ICS provides a mono-energetic beam of $\gamma$-rays which can be highly polarised but leads to a lower yield of pairs.

 \section{LUXE setup}     
\begin{figure}[htbp]
\centering
\includegraphics[width=0.99\textwidth]{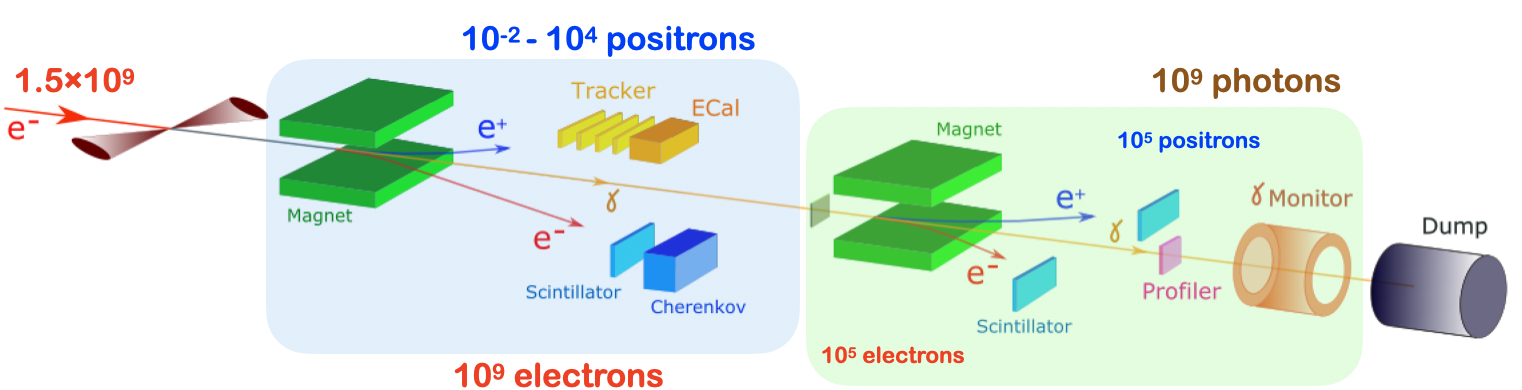}
\caption{Schematic experimental layout for the electron-laser setup. Two main detector subsystems are highlighted: Electron positron spectrometer - in blue and Photon detection system - in green. Shown are the magnets, detectors and absorbing elements.} 
\label{fig:layout} 
\end{figure}

 %To achieve the precision  measurements of strong field QED  in  the  transition  from  the  perturbative  to  the  non-perturbative  regime,  an  excellent performance of detection system is required in the range $\xi \sim 1 - 10 $.
 There are several qualitatively very different domains in LUXE and it is important that there is a sufficient redundancy in the measurements to ensure that systematic uncertainties can be estimated and minimised.
 The basic experimental concept of LUXE is shown for the case of electron-laser collisions in figure~\ref{fig:layout}. At the  interaction point (IP) each bunch containing $1.5 \cdot 10^9$ electrons  collides with a 800 nm laser pulse and final state particles are measured subsequently in a system of detectors. The LUXE setup conceptually contains two detector subsystem: Electron positron spectrometer (shaded in blue in figure~\ref{fig:layout}) and Photon detection system (PDS) - shaded in green. 
In the region behind the IP, the number of positrons per bunch crossing (BX) that needs to be measured ranges from $10^{-3}$ to $10^4$, on electron side the situation changes depending on the collision type: electron-laser or gamma-laser. The forward part of the setup intended for photon measurements covers a region of very high flux.
Due to this fact different subsystems and detector technologies are chosen for the various regions of the experimental setup. To study the performance all of the components of the experiment, the geometry model of the LUXE setup is implemented in GEANT4~\cite{geant4}.

\subsection{Positron detection}

The number of positrons per BX produced in the collisions is between $10^{-3}$ and $10^3$ in the initial phase of the experiment, and increases up to values of about $10^4$ for phase-1. 
%On the positron side, the expected rates for signal are $\sim 10^{-2} - 10^4$ depending on $\xi$ 
The detector has to provide a high efficiency for positrons reconstruction and good rejection of the background. In particular at low $\xi$, the signal rates are much lower than the background and thus detector technologies with a strong capability to reject the latter are required. The better the background rejection, the lower $\xi$ values can be measured and thus the fully perturbative regime of QED can be examined. The technology selected for the LUXE IP positron tracker is based on the ALPIDE silicon pixel sensor developed for the upgrade of the inner tracking system of the ALICE experiment at the LHC~\cite{Mager:2016Alpide}. 
 To ensure a reliable track reconstruction, the IP pixel tracker has four layers with a basic unit of the detector  - a $\sim 27 {\times} 1.5~\mathrm{cm^{2}}$ stave.  Two staves with overlap cover roughly 50 cm which are optimised for the magnet geometry and field.
The tracker has shown a detection efficiency above 98\% for the energy of positrons above 2 GeV,  a spatial resolution of around $\sim 5~\mathrm{\mu m}$,  and the energy resolution below 1\%. The sensor is able to tolerate an ionisation dose of up to 2.7 Mrad, which is sufficient for LUXE.

The tracker will be followed by an ultra-compact sampling electromagnetic calorimeter (ECal) based on the technologies developed for the ILC by the FCAL collaboration~\cite{Fcal_compact}. 
The ECal will allow to measure the energy spectrum of the positrons and to reject low energy background. 
It will also provide information for energy-position calibration. The calorimeter is designed as a sampling calorimeter composed of 20 layers of 3.5 mm ($1X_0$) thick tungsten absorber plates, with sensor planes placed in a 1 mm 
%$1 \thinmuskip \mathrm{mm}$
gap between absorber plates. 
For the sensor material two options are considered: Si or GaAs. The thickness of the silicon sensor is 320 $\mathrm{\mu m}$ and GaAs - 500 $\mathrm{\mu m}$; the pad size is $5 {\times 5}~\mathrm{mm^{2}}$. 
GaAs sensors are more radiation hard but were never used as sensitive layers in a sampling calorimeter. Prototypes of GaAs pad-sensors were investigated by the FCAL collaboration in a test beam and the response was found to be suitable for a calorimeter~\cite{testsGaAs}. Further test beam studies will follow.
The performance of the ECal was studied with GEANT4 simulations which showed an energy resolution of 19\% and a position resolution of 800 $\mathrm{\mu m}$.

\subsection{Electron detection}

The rate of electrons is substantially different in case of photon-laser and electron-laser collisions. LUXE uses different detectors for each of these experiment running modes.
In photon-laser mode the rate of electrons is the same as the rate of positrons, so the identical tracker will be used for their detection. It is installed on a movable platform and can be removed from the beam in electron-laser mode to prevent radiation damage. 

In case of electron-laser collisions, very high rates of electrons, up  to $10^9$ particles per BX, are expected, placing high demands on the radiation hardness and on the dynamic range of the detectors.
 In these region the signal is typically 10–100 times larger than the background.  Due to the high charged particle fluxes,  the detector technologies chosen for this region are gas Cherenkov detectors and  a scintillation screen. They provide complementary measurements of the electron spectrum. 
The scintillation screen offers a simple and inexpensive setup with a high position resolution. 
The large flux of electrons expected in the Compton process ($\sim 10^4 - 10^8 $) will induce enough light in a scintillating material to be imaged by a remote camera.
Similar technology is used by the AWAKE experiment that showed good performance for high flux with a linear response up to $\sim 2 \cdot 10^9$ electrons~\cite{BAUCHE2019103} .

The Cherenkov detector follows a design developed for polarimetry measurements at future lepton colliders, for which a two-channel prototype has been built and successfully operated in test-beam~\cite{Bartels:2010eb}. The Cherenkov device is not sensitive to low energy background electrons nor to photons.  
Simulations of the Cherenkov detector showed that the signal exceeds the small background by a factor of $10^{4}$. 

\subsection{Photon detection system}

\begin{figure}[ht]
    \begin{center}
    \includegraphics[width=0.6\textwidth]{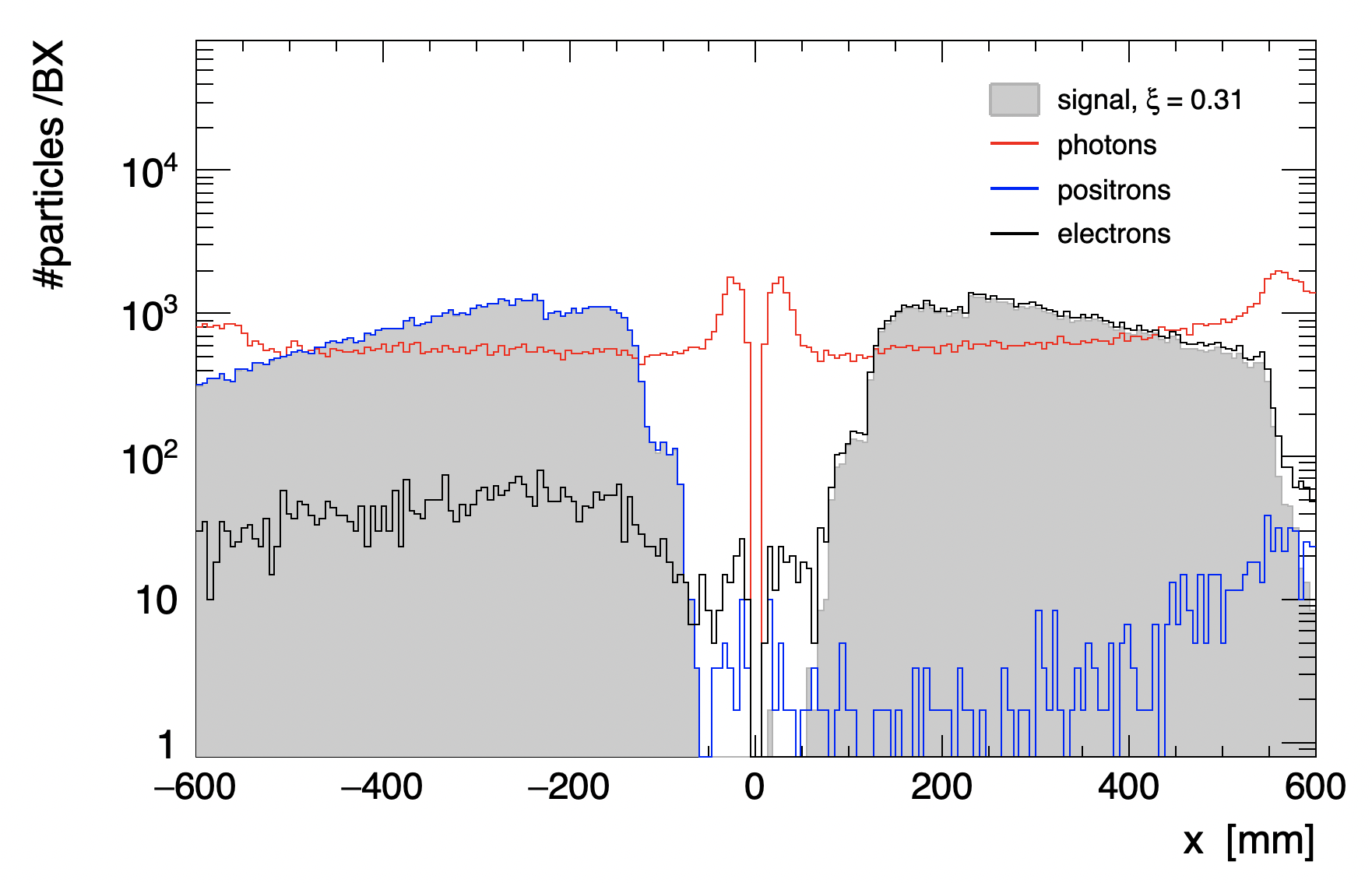}
    \caption{The number of particles per bunch crossing versus the x position in the scintillating screens of the forward photon spectrometer. The signal and background particles are estimated from the simulation of the e-laser setup. All particles intersecting the scintillating screens are considered regardless of whether their energy is above the screen's sensitivity threshold. The signal is shown for the JETI40 laser corresponding to $\xi$=0.31. 
}
      \label{fig:lanex_sig}
    \end{center}
\end{figure}

The photon detection system is located right downstream the electron
and positron spectrometer (figure~\ref{fig:layout}). The number of photons in non-linear Compton
scattering matches the number of electrons and reaches $10^9$. These photons are confined into a narrow cone with an aperture of about $1/\gamma {=} 1.7~\mathrm{mrad}$. 
For measuring their energy spectrum, the photons are partly converted into the electron-positron pairs in a thin target, then separated in another dipole magnet and registered in the two arms of the spectrometer detectors.
For a tungsten target of $10~\mathrm{\mu m}$ thickness the expected rate of electron-positron pairs is up to $10^5$  
and the same scintillator screens as in case of electron detection in the post-IP detector system are used to measure them. The photon energy spectra can be reconstructed via deconvolution of the measured spectra of the electrons and positrons with the Bethe-Heitler cross-section. Figure~\ref{fig:lanex_sig} shows the particle rates at the face of the scintillator screens on both arms of the spectrometer.
The first Compton edge is located around $x= \pm 100~\mathrm{mm}$ and is clearly
visible in the spectra of electrons and positrons. The step around $x= \pm 220~\mathrm{mm}$ is related to
the extension of the magnetic field in the lateral (x) direction. Electrons and
positrons with energies below 2.2 GeV exit the magnet laterally and thus
experience a shorter magnetic field. The number of background photons is
large, but they are rather inefficient in generating light in the
scintillator screens and thus do not affect the signal significantly.
The plot also illustrates that there is substantial leakage of
particles from the electron spectrometer located upstream. To reduce this
background, the shielding geometry and materials will be re-optimised in the future.
%The spectrometer will also be employed for monitoring the number of photons.   

\begin{figure}[ht!]
 \begin{center}
  \includegraphics[width=0.46\textwidth, height=4.6cm]{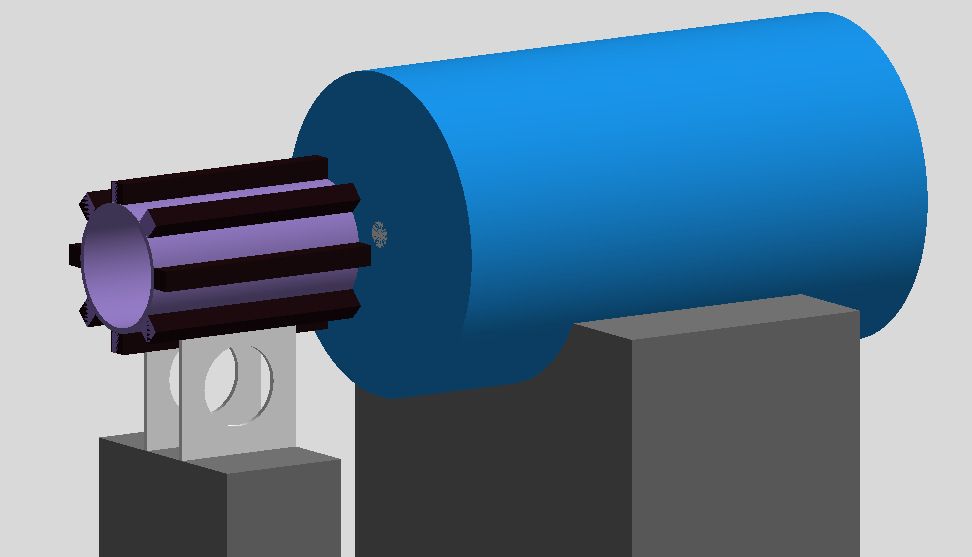}
    \includegraphics[width=0.45\textwidth]{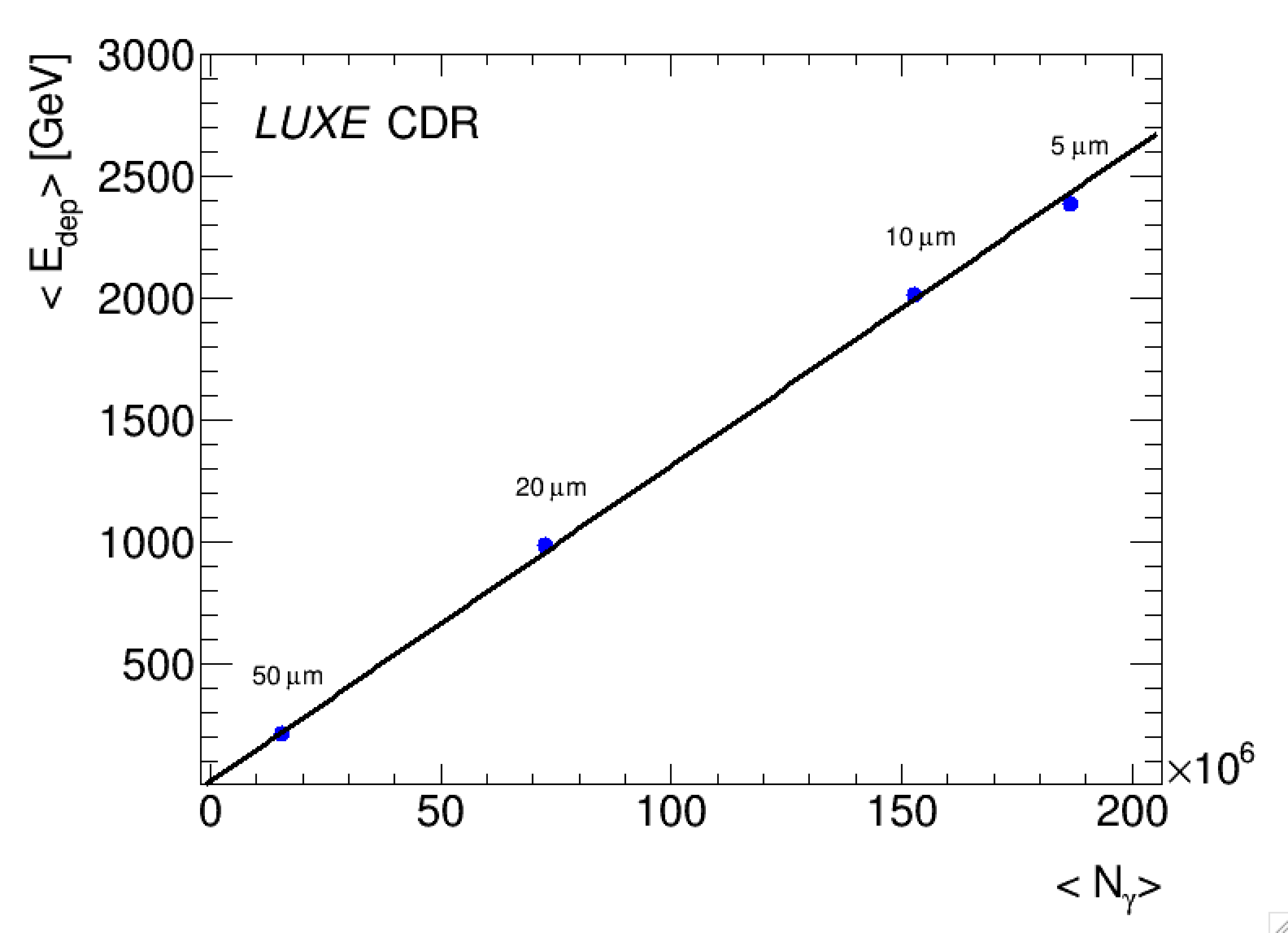}
      \end{center}
    \caption{(left) GEANT4 rendering for the backscattering calorimeter with the beam dump; (right) Average energy deposition versus the mean number of incident photons.}
    \label{fig:GM_perform}
\end{figure}

The high photon flux makes counting of photons rather challenging. A proposed solution is based on measuring the energy of the backscattered particles from the photon dump.
%to be sensitive to the direct photon flux variation. 
Such a detector - the backscattering calorimeter - provides a simple and robust way to monitor the variation of the photon flux with time and would play a role similar to a luminosity monitor in collider experiments. This detector is assembled from eight blocks of lead-glass oriented longitudinally with respect to the beam line, located $10~\mathrm{cm}$ upstream of the beam dump at a radius of $12~\mathrm{cm}$ around beam axis (see figure \ref{fig:GM_perform}, left). 
The performance of the backscattering calorimeter was studied in simulations and an almost linear dependence of the deposited energy on the number of incident photons was found (figure \ref{fig:GM_perform}, right). The design was optimised to reduce the radiation load to ensure a reasonable lifetime of the detector. The estimated uncertainty on the number of measured photons is 3-10\% depending on laser intensity.
 
To measure the angular spectrum of the photons, with a precision better than $0.5~\mathrm{\mu rad}$, the gamma ray profiler is placed behind the forward photon spectrometer. Photons in different polarization states are predicted to possess very different energies and angular distributions \cite{king2020}. For Compton scattering with polarised laser the gamma ray profiler will provide the measurement of the laser intensity with a precision better than 1\%. It is foreseen to be a set of two sapphire strip detectors with a pitch of 100 $\mathrm{\mu m}$ staggered behind each other with a relative rotation of $90^{\circ}$.  The extremely low leakage current ($\sim \mathrm{pA}$) at room temperature even after high dose irradiation makes sapphire detectors practically noiseless~\cite{Karacheban:sapphire}. The annual radiation dose is calculated as 10 MGy assuming a flux of $10^9$ photons per BX, $10^7$\,s of operation which (see Table~\ref{tab:dose}) corresponds approximately to the dose the detector can withstand.

\subsection{Radiation doses}

The ionisation doses for different detector subsystems were estimated by GEANT4 using the \textsc{QGSP\_BERT\_HP} physics list which includes physics processes for neutrons down to thermal energies. Table~\ref{tab:dose} summarises the average annual doses during one year of operation and the tolerable - for each detector type. Chosen detector technologies are within the limits to provide the measurements during the lifetime of the experiment except of the gamma ray profiler which is planned to be exchanged each year.
%Except the gamma ray profiler is it planned to be exchange the detectors each year
%corresponds approximately to the dose the detector can withstand, thus it would need to be replaced each year.
\begin{table}[htbp]
    \centering
        \caption{Annual average dose at the various detector locations for the e-laser setup based on GEANT4 simulations. The calculation assumes $10^7$\,s of operation. Also shown is the dose that can be tolerated by each detector type without performance degradation.}
    \begin{tabular}{|l|c|c|}
    \hline
         Detector &  avg. annual dose [Gy] & tolerable dose [Gy]\\\hline
         IP Pixel Tracker&  $10$ & $27\cdot 10^3$ \\
         IP Calorimeter ECal& $1$ & $10^6$\\
         IP Scintillator& $2\times 10^4$ & $10^8$ \\\hline
         PDS Scintillator & $10$ & $10^8$ \\
         PDS Gamma Profiler & $<10^7$ & $10^7$ \\
         PDS Backscattering Calorimeter& $0.5$ & $300$ \\
         \hline
    \end{tabular}
    \label{tab:dose}
\end{table}

\section{Summary}
The LUXE experiment will provide an exciting opportunity to explore QED in a new regime using the electron beam of the European XFEL and a high power laser. The scientific program includes the
observation of the transition from perturbative to non-perturbative regime of QED,
direct observation of pair production from the vacuum 
and accurate measurement of non-linear Compton scattering.
The experiment is planned to be installed in 2024 during an extended shutdown of the European XFEL.
%Recently released Conceptual design report~\cite{luxecdr} received positive DESY Physics Review Committee feedbacks with strong recommendation to proceed with Technical Design Report.
 The environment of the experiment is very inhomogeneous (with the range of particle rates of $10^{-3} - 10^9$) and is extremely challenging due to the high laser intensities that LUXE would explore. A dedicated detection system has been developed for the LUXE experiment employing silicon tracking, calorimetry, gas Cherenkov detectors as well as scintillating screens. It monitors the primary charged particle production as well as the photons with a dedicated system involving partial reconversion to electron-positron pairs as well as an instrumented dump. Despite the harsh radiation conditions experimental layout is well adapted to the physics tasks of the LUXE experiment.

\acknowledgments

This work has benefited from computing services provided by the German National Analysis Facility (NAF).
%This is the most common positions for acknowledgments. A macro is available to maintain the same layout and spelling of the heading.

%\paragraph{Note added.} This is also a good position for notes added after the paper has been written.


\end{document}